# Metaheuristic algorithm parameters selection for building an optimal hierarchical structure of a control system: a case study

1st Ruslan Zakirzyanov
*NEXT engineering, LLC*
Kazan, Russia
zr@nexteng.ru

*Abstract*—Metaheuristic algorithms are currently widely used to solve a variety of optimization problems across various industries. This article discusses the application of a metaheuristic algorithm to optimize the hierarchical architecture of an industrial distributed control system. The success of the algorithm depends largely on the choice of starting conditions and algorithm parameters. We examine the impact of parameter selection on the convergence of a modified ant colony algorithm and provide recommendations for tuning the algorithm to achieve optimal results for a specific industrial problem. The findings presented in this article can also be applied to other combinatorial optimization problems.

*Keywords—distributed control system, ant colony optimization, metaheuristics, hierarchical structure*

## I. INTRODUCTION

Complexity and a large number of variables that are controlled characterize modern industrial processes. To manage these processes, specialized control systems known as Distributed Control Systems (DCSs) [1] are utilized. These systems are hierarchical structures consisting of several levels. Each level contains various technical devices: sensors, actuators, input/output devices, controllers, network switches, servers, and operator workstations. Engineers face the complex task of designing an optimal DCS structure so that the system meets all specified requirements, ensures high-quality process control, and has a minimal cost. Since we can only use commercially available system components with factory-defined characteristics, the properties and functionality of such DCSs are largely determined by their structure. Designing an optimal structure is an important problem that is currently solved primarily empirically, relying on the designer's experience and recommendations from software and hardware manufacturers. Bioinspired metaheuristic algorithms are currently widely used to solve complex optimization problems. This article examines the application of a well-known Ant Colony Optimization (ACO) algorithm [2] to the problem of optimizing the structure of a distributed process control system. Based on the developed method, a series of computational experiments is implemented, which provides insight into how to tune the algorithm's parameters to achieve the best results.

## II. RELATED WORKS

The use of metaheuristic algorithms to solve optimization problems in various industrial fields is currently being actively developed by a large research community. These algorithms provide approximate solutions to problems, unlike exact methods such as Branch-and-bound (BnB) method or Integer Linear Programming (ILP), but can be effectively used to solve practical problems due to their lower resource intensity. The metaheuristic approach to solving optimization problems received explosive momentum after Fred Glover proposed the Tabu Search (TS) algorithm [3, 4]. The ACO algorithm was proposed by Marco Dorigo in 1992 to solve the Traveling Salesman Problem (TSP) [5]. Inspired by the behavior of ant colonies, this algorithm has become extremely widespread for solving a variety of problems, especially those represented as graphs. The number of diverse optimization algorithms based on the behavior of people, animals, or technical devices is currently very large, posing a new challenge for researchers: selecting the most appropriate algorithm for a specific problem.

Evaluating algorithm performance and comparing different algorithms for solving the same problem is also a pressing issue. For these purposes, researchers are developing specialized methods and benchmarks that allow for comparing the performance of various metaheuristic optimization algorithms and assessing the suitability of an algorithm for a specific task. The problem of optimizing the structure of an industrial system is discussed in [6]. The basic principles of representing a graph as a hierarchical structure are given in [7]. The application of reinforcement learning and the combination of various metaheuristic algorithms to improve their performance are discussed in [8]. Evaluating the performance of optimization algorithms and advanced evaluation methods using specialized benchmarks is described in [9, 10].

## III. PROBLEM STATEMENT

A detailed formal statement of the problem of constructing the optimal structure of a Distributed Control System is given in [11], [12]. We call this problem DCSSP (Distributed Control System Structure Problem).

In short, we represent the DCS structure as a tree (acyclic graph) $\mathcal{G} = (\mathcal{V}, \mathcal{E})$, where $v \in \mathcal{V}$ are devices (graph nodes) and $\mathcal{E}$ are graph edges (communication channels between devices). An example of such a tree is shown in Fig. 1. The number of levels in the structure is specified by the designer and can vary.

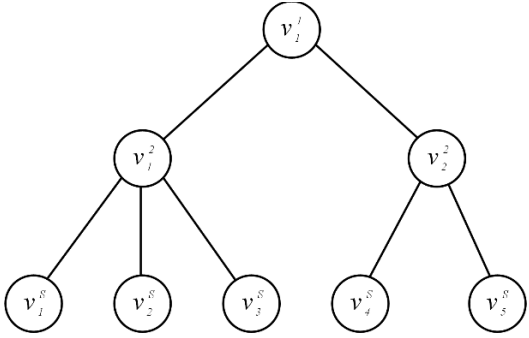

Fig. 1. Hierarchical structure of the DCS

The set of device types $\mathcal{U} = \{u_1, \ldots, u_U\}$, consisting of $U$ device types, is given.

Each node of the structure (device) of any type performs the same type of actions, consisting of three phases of the operating cycle of the device's internal program: reading data, processing data (implementation of control algorithms), and writing data. All the devices are divided into two groups: processors and repeaters.

Each type $u_i \in \mathcal{U}$ of structure node is characterized by the following parameters:

- Cost of the device $C_i \in \mathbb{R}^+$.
- Number of connected physical channels $N_i \in \mathbb{N}$.
- Maximum memory capacity $R_i \in \mathbb{R}^+$.
- Probability of device failure $P_i \in [0,1]$.
- Performance (execution time of one program instruction) $T_i \in \mathbb{R}^+$.
- Operating mode $y_i \in \{0,1\}$ (1 – processor, 2 – repeater).
- Maximum number of child devices $M_i \in \mathbb{N}$.
- Transmission delay for repeaters $\tau_i \in \mathbb{R}^+$.

Each device type is represented by a vector:

$$u_i = (C_i, N_i, R_i, P_i, T_i, y_i, M_i, \tau_i), \quad i = \overline{1, U}. \quad (1)$$

A set of control loops $\mathcal{A} = \{a_1, \ldots, a_A\}$, is also given. It consists of $A$ loops $a_j \in \mathcal{A}$:

$$a_j = (n_j, r_j, w_j), \quad j = \overline{1, J}, \quad (2)$$

where $n_j \in \mathbb{N}$ is the number of physical signals in the circuit, $r_j \in \mathbb{R}^+$ is the amount of memory required to store all instructions and variables of the circuit, $w_j \in \mathbb{N}$ is the number of instructions in the signal processing program of this circuit.

In addition, in [11], we introduce a number of auxiliary functions that help us describe the structure of the DCS in the form of a graph as close as possible to a real industrial system.

Also, we introduce additional variables: $x_{va} \in \{0,1\}$ and $z_{va} \in \{0,1\}$ [11]. If the control loop is physically connected to a node of leaf level, we will say that loop $a$ is connected to leaf $v$ ($x_{va} = 1$). If the signals of loop $a$ are processed in node $v$, we will say that loop $a$ is assigned to node $v$ ($z_{va} = 1$).

We also introduce a large number of restrictions into the system, including limitations on system reliability, performance, the number of child devices, etc.

The optimal hierarchical structure $\mathcal{G}_O$ should minimize the cost of the system under the specified constraints. Thus, the optimal structure can be determined as a result of solving the optimization problem:

$$C_O = \min_{\mathcal{G}} \sum_{v \in \mathcal{V}} C_v, \quad (3)$$

where $C_O$ is the optimal total cost of creating and operating the system.

Thus, the structural optimization problem is formulated as follows. It is necessary to select such a hierarchical structure from nodes of predetermined types in order to deliver the minimum of the objective function (3) under the given constraints. We assume that an unlimited number of devices of each type can be used. We also assume that the desired structure is a Proper Hierarchy Graph [7].

IV. APPLICATION OF THE METHOD

The ACO algorithm was chosen to solve the DCSSP. This algorithm is currently widely used to solve various optimization problems, especially those represented as graphs. A detailed description of the ant colony algorithm is given in [5]. Unlike other metaheuristics (e. g. GA, TS, GWO), it was designed to solve problems on a graph and can find solutions to problems with multiple constraints where other metaheuristics fail. A distinctive feature of the algorithm is that it remembers the best paths using pheromones, which evaporate during subsequent iterations of the algorithm. Each step is selected according to the following rule:

$$P_i = \frac{\tau_i^\alpha \eta_i^\beta}{\sum_{k=0}^{N} \tau_k^\alpha \eta_k^\beta}, \quad (4)$$

where $P_i$ – probability of choosing the $i$-th option, $\tau$ – pheromone quantity, $\eta$ – option cost parameter, $\alpha$ – pheromone weight, $\beta$ – heuristic weight, N – the total number of acceptable options.

The pheromone quantity is updated according to the formula:

$$\tau_{i+1} = (1 - \rho)\tau_i + \Delta\tau_i, \quad (5)$$

where $\rho$ – evaporation coefficient of the pheromone, $\Delta\tau_i$ – the amount of deposited pheromone.

To solve the DCSSP problem, we divided each ant's step into several actions (selecting the type of new device, selecting the number of child devices, assigning loops, etc.). Each choice is performed with a probability determined by expression (4).

The ant sequentially builds a tree from the root to the leaves by adding graph vertices. Each iteration of the algorithm terminates if a feasible tree is constructed. After construction, constraints are checked. If the constraints are violated, the solution is discarded.

The metaheuristic algorithm used does not provide an exact solution. The proximity of the obtained solution to the optimal one is largely determined by the method for specifying the initial conditions and the choice of algorithm parameters. Furthermore, the algorithm tends to get stuck on local optima. A fairly large number of algorithm modifications have been developed to minimize the probability of getting stuck on local optima [12]. We supplemented the ACO with an algorithm for Local Search (LS) near the expected optimum. The LS algorithm used the 2-opt swap for devices of different types after the structure solution was found. The results of a computational experiment showed that the modified algorithm is quite effective in avoiding local optima.

To implement the selected algorithm, specialized software was developed. We used Python and Qt for development. The program has a GUI and allows for setting device and control loop parameters, the number of hierarchy levels, and quantitative system characteristics. The software also allows for setting ACO algorithm parameters, such as the number of iterations, the number of ants, $\alpha$, $\beta$, and $\rho$ parameters. Notably, the developed software allows for specifying algorithm parameters not only as constants but also as functions whose argument is the current iteration number.

For the first computational experiment, we introduced two device types into the hierarchy. The $u_1$ processor has characteristics similar to those of a Programmable Logic Controller (PLC), widely used in industry. A PLC is quite expensive but lacks I/O channels. The $u_2$ repeater has characteristics similar to an I/O module, which can connect up to eight physical signals. Both device types have four network ports on the board for connecting downstream devices.

For all the computational experiments implementation, a hardware platform with the following characteristics was specially allocated: Intel Core i5 Processor, 16 GB RAM, 512 GB SSD.

A graphical representation of the optimal structure for values of the number of loops $A = 50$ and the number of hierarchy levels $S = 3$ is shown in Fig. 2. The number of connected control loops is indicated at the bottom. The convergence graph of the algorithm is shown in Fig. 3. The Y axis shows the cost of the system, and the X axis shows the number of algorithm iterations. We show the average (blue) and the best (red) solution due to 30 algorithm runs.

The choice of algorithm parameters significantly affects the convergence. Using the developed software, a number of test examples were solved to evaluate the influence of the parameters on the convergence of the algorithm. Calculations were performed for a system with four hierarchy levels ($S = 4$), the number of device types $U = 5$, and the number of control loops $A = 200$. The parameters specified in Table 1 were selected for the experiment. The developed software allows specifying the algorithm parameters both as constants and as functions of the argument $n$, where $n$ is the algorithm iteration number. Based on the data obtained as a result of the computational experiments, algorithm convergence graphs were constructed. The graphs are shown in Fig. 4. As before the Y axis shows the cost of the system, the X axis shows the number of algorithm iterations.

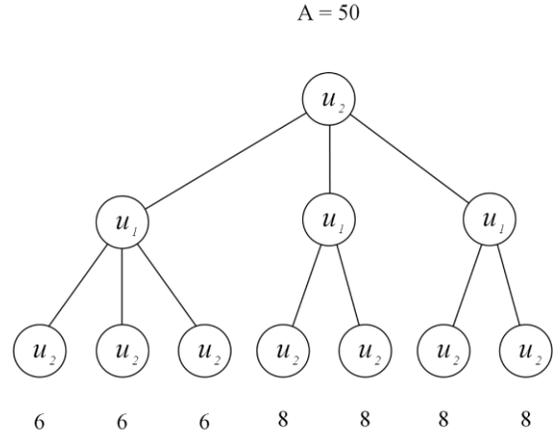

Fig. 2. Optimal structure of DCS

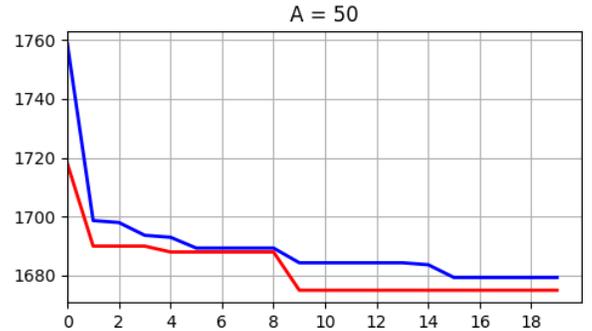

Fig. 3. Convergence graph of the algorithm

TABLE I. ALGORITHM PARAMETERS

| Exp. No | Algorithm Parameters | | |
|---|---|---|---|
| | $\rho$ | $\alpha$ | $\beta$ |
| 1 | 0.25 | 2.0 | 1.0 |
| 2 | 0.25 | 2.0 | 0.0 |
| 3 | 0.25 | $2/(n + 0.01)$ | $0.1n$ |
| 4 | 0.25 | $0.2n$ | $1/(n + 0.01)$ |

In Fig. 4, we show the average convergence for the results from Table 2 and the best solution (red curve) due to 30 algorithm runs as before. Experimental results are shown in Table 2. We provide average and best solutions for each experiment, and we compute the Coefficient of variation (CV) as well.

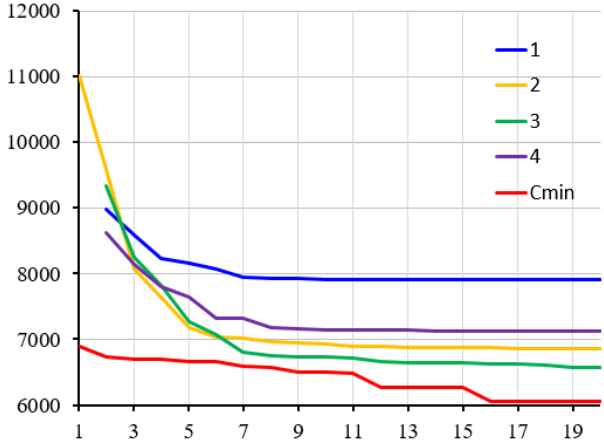

Fig. 4. Convergence graph of the algorithm for different algorithm parameters

TABLE II. EXPERIMENTAL RESULTS

| Exp. No | Algorithm Parameters | | |
|---|---|---|---|
| | $C_{min}$ | $C_{avg}$ | CV, % |
| 1 | 6458.00 | 7903.20 | 13,59 |
| 2 | 6286.00 | 6848.10 | 3.95 |
| 3 | 6063.00 | 6567.20 | 3.39 |
| 4 | 6455.00 | 7122.50 | 12.74 |

As we can see, strategy No. 3 to decrease $\alpha$ and to increase $\beta$ produces the best performance. We can recommend using these results for the practice implementation and further research.

## V. FURTHER RESEARCH

This problem offers significant potential for continued and expanded research. An important area of research is the study of the possibility of using other metaheuristic algorithms (GA, GWO, TS) to solve this problem and a quantitative comparison of their effectiveness. Various modifications of the ACO algorithm are also being considered to improve its efficiency for larger problem dimensions. Fine-tuning the algorithm parameters to most effectively solve a specific practical problem is also a significant topic for research.

It should also be noted that certain assumptions are made in the problem conditions (such as the absence of horizontal connections between hierarchy nodes and so on). Taking these assumptions into account in the model may become the subject of additional research, development, and generalization of the results of this article, but already now we can provide basic solutions for evaluating the optimality of the structure.

The quality of the resulting control system can be verified by simulating the system on a specific control object, for example, a process unit at a chemical plant.

## VI. CONCLUSION

The DCSSP problem is currently extremely relevant, as in practice, the synthesis of distributed control system structures for continuous technological processes is carried out primarily empirically, based on the designer's experience and equipment manufacturer recommendations, and is not always optimal. The approaches proposed in this and related works can be used in the design of large-scale systems and also provide grounds for new research in this area.

REFERENCES

[1] M. Tkáčik, J. Jadlovský, S. Jadlovská, A. Jadlovská, T. Tkáčik "Modeling and Analysis of Distributed Control Systems: Proposal of a Methodology", Processes, 2024, 12, 5. doi: 10.3390/pr12010005.

[2] M. Dorigo, T. Stützle, "Ant Colony Optimization", MIT Press, 2004. doi: 10.7551/mitpress/1290.001.0001.

[3] F. Glover, V. Campos, R. Martí "Tabu search tutorial. A Graph Drawing Application", 2021, TOP 29, pp. 319–350. doi: 10.1007/s11750-021-00605-1.

[4] K. Sörensen, M. Sevaux, F. Glover "A history of metaheuristics", 2018. doi: 10.1007/978-3-319-07124-4_4.

[5] A. Colorni, M. Dorigo, V. Maniezzo. "Distributed optimization by ant colonies", Proceedings of the First European Conference on Artifcial Life, MIT Press, 1992, pp. 134–142.

[6] F. Sarı, S. Nigdeli, G. Bekdas, Z. W. Geem "A Review of Optimization of Structural Control Systems: Usage of Metaheuristics in Structural Control", 2024. doi: 10.4018/979-8-3693-2161-4.ch007.

[7] S. Cavero, E. G. Pardo, F. Glover, R. Martí "Strategic oscillation tabu search for improved hierarchical graph drawing", Expert Systems with Applications, 2024, vol. 243. doi: 10.1016/j.eswa.2023.122668.

[8] O. A. Arık, G. Toga, B. "Atalay Reinforcement Learning Based Acceptance Criteria for Metaheuristic Algorithms", International Journal of Computational Intelligence Systems, 2025, vol. 18, No. 207. doi: 10.1007/s44196-025-00924-2.

[9] S. M. Almufti and A. A. Shaban, "Comparative Analysis of Benchmark Optimisation and Real-World Applications," FMDB Transactions on Sustainable Energy Sequence, vol. 3, no. 1, pp. 51–61, 2025, doi: 10.69888/FTSES.2025.000417.

[10] S. M. Almufti, R. R. Asaad, A. A. Shaban, R. B. Marqas "Benchmarking metaheuristic algorithms: a comprehensive review of test functions, real-world problems, and evaluation metrics", Journal of Electronics, Computer Networking and Applied Mathematics, 2025, No 5(2), pp. 16-35. doi: 10.55529/jecnam.52.16.35.

[11] R. Zakirzyanov, "Application of Metaheuristic Algorithms for Optimizing the Structure of Industrial Control System," 2025 VI International Conference on Control in Technical Systems (CTS), Saint Petersburg, Russian Federation, 2025, pp. 34–36, doi: 10.1109/CTS67336.2025.11196304.

[12] R. Zakirzyanov, "Structural Optimization of Software and Hardware Complex of Automated Process Control Systems of Oil and Gas Industry Enterprises Using Metaheuristic Algorithms," 2025 International Russian Automation Conference (RusAutoCon), Sochi, Russian Federation, 2025, pp. 785–789, doi: 10.1109/RusAutoCon65989.2025.11177422.